\documentstyle[emulateapj,epsf]{article}

\lefthead{Lorimer \& Xilouris}
\righthead{PSR~J1907+0918}

\begin{document}

\slugcomment{Accepted for publication in the Astrophysical Journal}

\title{PSR~J1907+0918 --- a young radio pulsar near SGR~1900+14 and G42.8+0.6}
\author{Duncan R. Lorimer}
\affil{Arecibo Observatory, HC3 Box 53995, Arecibo, Puerto Rico, PR 00612}
\author{Kiriaki M. Xilouris}
\affil{Department of Astronomy, University of Virginia, Charlottesville,
VA 22903}

\begin{abstract}
\noindent
We have extensively searched for periodic signals from the soft-gamma
repeater SGR~1900+14, at 430 and 1410 MHz with the Arecibo
telescope. Our observations did not reveal the 5.16-s periodicity
discovered at X-ray wavelengths by Hurley et al.~(1998). We place
pulsed flux-density upper limits of 150 and 30 $\mu$Jy at 430 and 1410
MHz respectively.  In the course of the 1410-MHz search we discovered
a 226-ms radio pulsar, PSR~J1907+0918. Its period derivative implies
that the age of J1907+0918 is only 38 kyr, making it one of the
youngest members of the known pulsar population. Independent lines of
evidence in support of this apparent youth are the unusually high
degree of circular polarization and a relatively flat radio
spectrum. The close proximity of this young radio pulsar to the
supernova remnant G42.8+0.6 poses a problem for the proposed
association between the G42.8+0.6 and SGR~1900+14.
\end{abstract}

\keywords{stars: neutron --- pulsars: individual (PSR J1907+0918) ---
stars: individual (SGR 1900+14) --- supernova remnants: individual (G42.8+0.6)}

\section{Introduction}
\label{sec:intro}

Since the discovery of the first soft-gamma repeater (SGR), the source
of the famous ``March 5'' burst from the supernova remnant N49 in the
Large Magellanic Cloud \nocite{mgi+79,cdt+82} (Mazets et al.~1979;
Cline et al.~1982), three other SGRs have been found close to the
Galactic plane \nocite{mgg79,lff+86,kkw+98} (1900+14: Mazets,
Golenetskii \& Gur'yan 1979; 1806--20: Laros et al.~1986; 1627--41:
Kouveliotou et al.~1998b). Recently, Cline et al.~(2000)
\nocite{cfg+00} report a possible fifth SGR, 1801--23. Several lines
of evidence support the notion that SGRs are magnetars --- young,
slowly-rotating neutron stars with intense ($\sim 10^{14}$ G) magnetic
fields \nocite{dt92a,td95,td96b} (Duncan \& Thompson 1992; Thompson \&
Duncan 1995,1996). Firstly, all SGRs presently known are located close
to or within a supernova remnant. SGR~1900+14 lies just outside the
remnant G42.8+0.6 \nocite{vkfg94,hkp+99} (Vasisht et al.~1994; Hurley
et al.~1999a). Secondly, coherent pulsations in the range 5--8 s have
been observed in three SGRs \nocite{kds+98,hkms98} (Mazets et
al.~1979; Kouveliotou et al.~1998a; Hurley et al.~1998).  Finally, the
measurement of high spin-down rates in two SGRs \nocite{ksh+99,mrl99}
(Kouveliotou et al.~1998a,1999; Marsden, Rothschild \& Lingenfelter
1999) are suggestive of a neutron star spinning down via relativistic
wind emission.

Following the precise localization and discovery of coherent 5.16-s
X-ray pulsations from SGR~1900+14 during an intense outburst in 1998
\nocite{hkms98} (Hurley et al.~1998), we undertook a high-sensitivity
search for radio pulsations from SGR~1900+14 using the upgraded
Arecibo telescope. In this {\it Letter}, we report on these
observations and discuss their implications.  The pulsar search is
described in \S \ref{sec:search}. Although we did not detect the
5.16-s periodicity, we did find a young radio pulsar, J1907+0918,
located 2 arcmin from SGR~1900+14. In \S \ref{sec:1907} we summarize
15 months of Arecibo timing observations of PSR~J1907+0918. As we
discuss in \S \ref{sec:disc}, our discovery of PSR~J1907+0918 is the
latest in a number of independent pieces of evidence which challenges
the proposed association between SGR~1900+14 and G42.8+0.6.

\section{Search for radio pulsations from SGR~1900+14}
\label{sec:search}

In early June 1998 we observed SGR~1900+14 using the Arecibo telescope
seven days after the source became active following a long period of
quiescence. Coordinated observations of the outburst by Ulysses and
BATSE/CGRO resulted in an accurate localization of the source position
($\alpha_{2000}=19^{\rm h}07^{\rm m}15\fs34$;
$\delta_{2000}=+09^\circ19'21\farcs7$) kindly provided to us in
advance of publication by C.~Kouveliotou. This allowed us to make deep
search observations.  The Arecibo search for radio pulsations was
carried out with the Penn State Pulsar Machine (PSPM), a filterbank
which records the total-power outputs of the receiver over
$128\times60$-kHz frequency channels every 80 $\mu$s. We collected
continuous PSPM data in search mode for 1380 s using the 430-MHz line
feed (June 10) and the Gregorian L-narrow receiver at 1410 MHz (June
11). The error in the localized position (a 3 arcmin$^2$ error
ellipse) was sufficiently small to require only a single telescope
pointing at each of the two frequencies (the 3-dB width of the Arecibo
beam is 10 arcmin at 430 MHz and 3 arcmin at 1410 MHz).

These data were searched for the presence of periodic signals using
software developed for a recent pulsar survey with the Effelsberg
radio telescope \nocite{lkm+00} (Lorimer et al.~2000). The search
explores the three-dimensional parameter space defined by pulse
period, pulse duty cycle and dispersion measure by de-dispersing and
Fourier transforming the raw data before searching for statistically
significant harmonic features in the amplitude spectra. The best
candidates are folded in the time domain to produce diagnostic plots
which are saved for later inspection.

To estimate the sensitivity of these searches, following \nocite{dss+84}
Dewey et al.~(1984), we calculate the minimum flux density $S_{\rm min}$ 
required for a detection as follows:
\begin{equation}
\label{equ:smin}
   S_{\rm min} = \frac{\beta \, \, \sigma_{\rm min} \,
   (T_{\rm rec}+T_{\rm sky})}
  {G \, \sqrt{N_p \, \Delta \nu \, t_{\rm int}}} \sqrt{\frac{W}{P-W}}.
\end{equation}
Here the factor $\beta \simeq 1.1$ reflects losses due to hardware
limitations, $\sigma_{\rm min}=8$ is the threshold signal-to-noise
ratio, $T_{\rm rec}$ and $T_{\rm sky}$ are the receiver and sky noise
temperatures (K), $G$ is the effective antenna gain (K Jy$^{-1}$),
$N_p=2$ is the number of polarizations summed, $\Delta
\nu=128\times60$ kHz $=7.68$ MHz is the total observing bandwidth,
$t_{\rm int}=1380$ s is the integration time, $W$ is the detected
pulse width and $P$ is the pulse period.  For our 430-MHz search we
can insert the above values along with $T_{\rm rec}=100$ K, $T_{\rm
sky}=150$ K, $G=10$ K Jy$^{-1}$ to find $S_{\rm min,430} \simeq 1.5
(W/P)^{1/2}$ mJy.  For the 1410-MHz search, $T_{\rm rec}=35$ K,
$T_{\rm sky}\simeq7$ K, $G=8$ K Jy$^{-1}$ so that $S_{\rm min,1410}
\simeq 0.3 (W/P)^{1/2}$ mJy.  In both these cases, we have assumed
that $W \ll P$.  Note also that both the 430-MHz and 1410-MHz
observations were carried out at necessarily high zenith angles
($>10^{\circ}$) so that the quoted antenna gains are lower than those
applicable to observations closer to the zenith (see
http://www.naic.edu/aomenu.htm for up-to-date telescope information).

These search observations were carried out as part of a larger search
for pulsars in supernova remnants at Arecibo. As part of this project,
we undertook a number of calibration observations with the PSPM for
pulsars with well-known flux densities and pulse widths. The
signal-to-noise ratios predicted from equation \ref{equ:smin} compare
well with those obtained in practice and give us confidence in the
above flux limits obtained from our blind search.

Our search was carried out before the announcement of the 5.16-s
periodicity by Hurley et al.~(1998). Following this discovery, and
Shitov's (1999) \nocite{shi99} detection at 111 MHz, we de-dispersed
and folded both sets of search data at the predicted topocentric
period to look for evidence of a pulsed signal. We found no convincing
evidence for pulsar-like profiles above a signal-to-noise threshold of
6. This effectively reduces the above limits from the blind
periodicity searches by a factor of 6/8 to $S_{\rm min,430} \simeq 1.1
(W/P)^{1/2}$ mJy and $S_{\rm min,1410} \simeq 0.2 (W/P)^{1/2}$ mJy.
Shitov, Pugachev \& Kutuzov (2000) \nocite{spk00} report a pulse width
of 100 ms for the 5.16-s radio pulsations from SGR~1900+14 observed at
111 MHz. Based on the above sensitivity estimations, such a pulsed
signal would be detectable down to a flux-density limit of 150 $\mu$Jy
at 430 MHz and 30 $\mu$Jy at 1410 MHz.  We note that these upper
limits, together with Shitov et al.'s flux measurement of SGR~1900+14
as a 50-mJy radio pulsar at 111 MHz, constrain the power-law index of
the radio spectrum to be steeper than --3. Clearly, further radio
observations at lower frequencies are required to confirm or refute
the 111-MHz detection reported by Shitov et.~al.

\section{Discovery and follow-up of PSR~J1907+0918}
\label{sec:1907}

During the analysis of our 1410-MHz search observations towards
SGR~1900+14 we found a very promising 113-ms pulsar candidate.
Subsequent 1410-MHz observations made in September 1998 confirmed the
existence of the pulsar (J1907+0918) and identified its true period to
be 226 ms \nocite{xkl+98} (Xilouris et al.~1998).

In order to accurately determine the spin and astrometric properties
of PSR~J1907+0918, we have been carrying out regular timing
measurements using the PSPM since October 1998. A preliminary
ephemeris, based on the discovery and confirmation observations, was
used to predict the apparent pulse period initially. This period was
sufficiently accurate to be used by the PSPM in timing mode to fold
the incoming data from each of the 128 frequency channels for
typically 90 s before saving the profiles to disk. The individual
frequency channels were later de-dispersed to produce a time-tagged
integrated pulse profile for each 90-s observation. For each of these
scans, the mean pulse time of arrival (TOA) at the observatory was
then determined by cross correlating the profile with a high
signal-to-noise template (for further details, see Taylor
\nocite{tay92} 1992).

The {\sc tempo} software package (which is freely available at
http://pulsar.princeton.edu/tempo) and the DE200 planetary
ephemeris were then used to correct these topocentric TOAs for the
Earth's motion around the Sun and transform them to the frame of the
solar system barycenter before fitting them to a simple isolated
pulsar spin-down model \nocite{mt77} (see e.g.~Manchester \& Taylor
1977).  Multiple passes of the TOAs through {\sc tempo} were required
in order to minimize the model minus observed timing residuals and resolve
any pulse numbering ambiguities. The resulting ephemeris based on
observations spanning a 15-month baseline between October
1998 and January 2000 yields a sub-arcsecond position and highly
accurate spin-down parameters which are presented in Table
\ref{tab:1907}. The post-fit timing residuals for this ephemeris
are free from systematic trends at the level of 108 $\mu$s rms.

\medskip
\epsfxsize=7truecm
\epsfbox{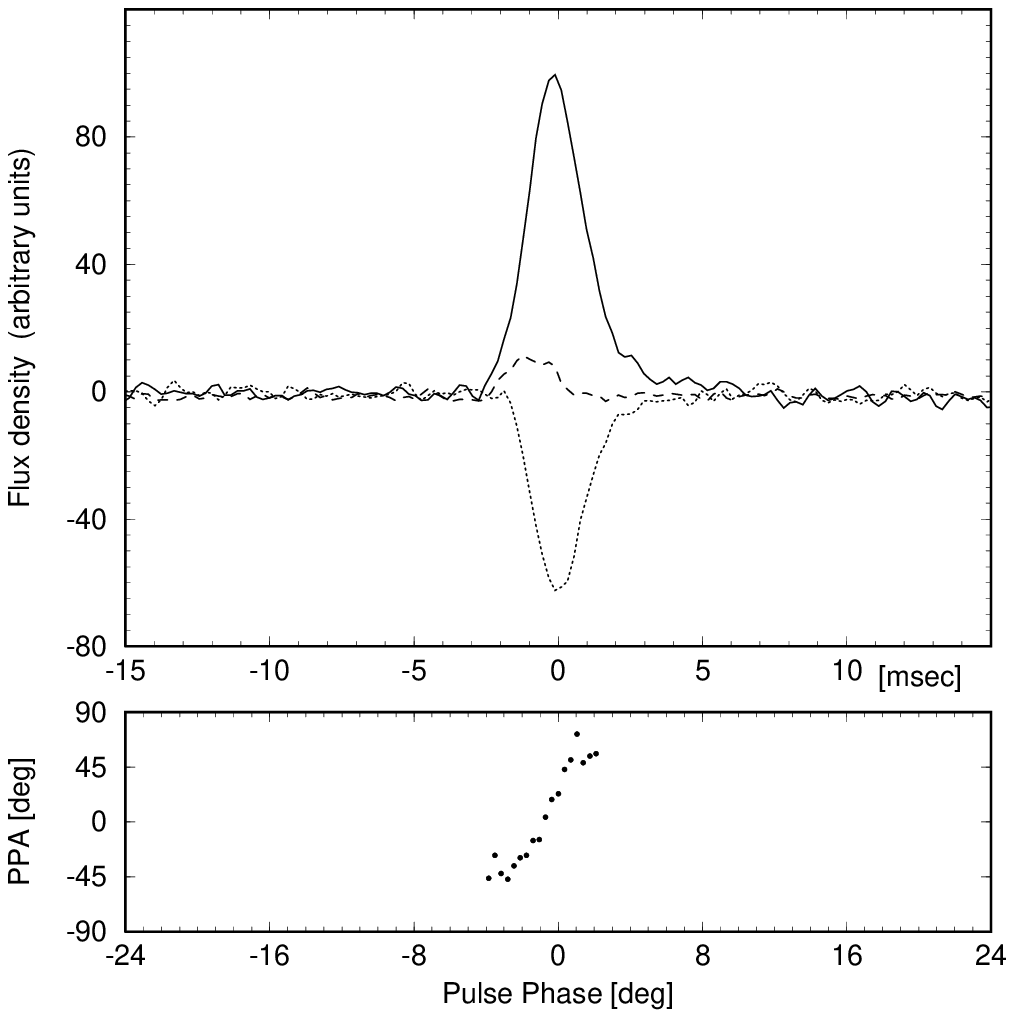}
\figcaption{\label{fig:pprof} Polarization profile of PSR~J1907+0918
at 1410 MHz. In the upper panel the total power (solid line), linearly
polarized power (dashed line), and circularly polarized power (dotted
line) are shown. The lower panel shows the variation of the linear
polarization position angle (PPA) with pulse phase.  Note the scale of
the abscissa is such that this profile represents only 13\% of the
entire pulse phase. This profile is freely available at the EPN data
archive http://www.mpifr-bonn.mpg.de/div/pulsar/data}
\bigskip

From the spin parameters of PSR~J1907+0918, we infer
a dipole surface magnetic field strength of $4.7\times10^{12}$ G and a
characteristic age of only 38 kyr. This low characteristic age places
J1907+0918 in the youngest 3\% of known Galactic radio pulsars. Many
pulsars with similar characteristic ages have
relatively flat radio spectra (Johnston 1990; \nocite{joh90} Lorimer
et al. \nocite{lylg95} 1995) and are strongly polarized sources
at radio frequencies above 1 GHz \nocite{hoe99} (von Hoensbroech 1999). 

In order to investigate the emission properties of PSR~J1907+0918, in October
and November 1998 we carried out a series of quasi-simultaneous
multi-frequency observations at Arecibo using the 430-MHz line feed,
along with the Gregorian receivers centered at 1410, 2380 and 5000
MHz. We used the Arecibo-Berkeley Pulsar Processor (ABPP) for
these observations. The ABPP is a 32-channel coherent-dispersion-removal 
machine capable of high-precision timing and polarimetry (for details see 
\nocite{bdz+97} Backer et al.~1997 and http://www.naic.edu/$^{\sim}$abpp)
which allowed us to carry out high-quality
observations spanning a bandwidth of up to 35 MHz.

In Fig.~\ref{fig:pprof} we present a high signal-to-noise polarization
profile of PSR~J1907+0918 based on ABPP observations carried out at
1410 MHz.  The degree of circular polarization in this profile is
58\%, making this the pulsar with the highest known degree of circular
polarization at 1410 MHz. PSR~J1907+0918 is ideally suited as a calibrator 
for future pulsar polarimetry observations. 
As can be seen from the scale in Fig.~\ref{fig:pprof}, the pulse duty
cycle of J1907+0918 is extremely small, less than 2\% at observing
frequencies above 1 GHz. Fewer than 2\% of all known pulsars have such
a narrow pulse. At 430 MHz, however, the lowest frequency that we have
data for, the profile appears to be broadened due to multi-path scattering. 
Fitting the profile to a truncated exponential function,
we determine a scattering time-scale at 430-MHz of 17 ms. 
The flux densities $S_{\nu}$ at each frequency $\nu$ are summarized 
in Table 1. Fitting these to a power law of the form $S_{\nu} \propto
\nu^{\alpha}$ yields a mean spectral index $\alpha=-0.3\pm0.2$. Only
two pulsars out of 260 listed by Lorimer et al.~(1995) have flatter
spectra.  In summary, based on the timing and emission properties of
PSR~J1907+0918, we conclude that it is most likely a young radio pulsar.

\section{Discussion}
\label{sec:disc}

Although our radio search for the 5.16-s pulsations from SGR~1900+14 was
unsuccessful, the discovery of PSR~J1907+0918 has important
implications for the origins of SGR~1900+14 and its proposed 
association with the supernova remnant G42.8+0.6. 

\subsection{PSR~J1907+0918 and SGR~1900+14}

The angular separation between PSR~J1907+0918 and SGR~1900+14 is $\sim
2$ arcmin. Apart from globular cluster pulsars and double neutron star
binaries, this is the closest pair of neutron stars on the plane of
the sky. Either the two stars are physically close to each other (at
the dispersion-measure distance of J1907+0918, 7.7 kpc, the projected
separation is only 5 pc), or they are at different distances and only
appear close when seen in projection.

To investigate the case of a simple geometric projection, we require
an estimate of the expected number of pulsars per unit area that would
be detectable in our deep search. Cordes \& Chernoff (1997) provide a
number of useful analytic expressions for these purposes. Starting
from their equation (8) we find the mean search volume $\overline{V}$
to be proportional to $S_{\rm min}^{-3/2}$. Here for simplicity we
have neglected any strong period dependence upon the sensitivity
(reasonable for long-period pulsars) and weighted the mean over a
pulsar luminosity function of the form $d\log N/d\log L=-1$
\nocite{lbdh93} (see e.g.~Lorimer et al.~1993). Under the simplest
assumption that the number density of pulsars is approximately
constant, the expected number of pulsars $N \propto \overline{V}
\propto S_{\rm min}^{-3/2}$.  To estimate the pulsar detection rate in
our search we extrapolate the results of the current Parkes multibeam
survey of the Galactic plane which is detecting about 1.5 pulsars
deg$^{-2}$ within $|b|<1^{\circ}$ around the location of PSR
J1907+0918 on the Galactic plane (F.~Camilo, private
communication). Under the above assumptions, we would expect a deep
Arecibo search of this part of the plane to detect around
$1.5\times2.5^{3/2}\simeq6$ pulsars deg$^{-2}$. Since the width of the
1410-MHz beam is around 3.5 arcmin, this corresponds to about one
pulsar detection every 60 telescope pointings. It is therefore not
unreasonable to appeal to simple chance detection in this case where
we have only made one pointing at 1410 MHz.

Aside from simple probability calculations, there is one other reason
to believe that the apparent close angular proximity between
PSR~J1907+0918 and SGR~1900+14 is simply a projection effect. If we
were to assume that the two neutron stars had a common binary origin,
this would require a binary system in which both stars are
sufficiently massive to undergo a supernova explosion. Although
theoretical arguments can be made to explain the close proximity of
PSR~B1853+01 and PSR~B1854+00 \nocite{wcd91} (Wolzczan Cordes \& Dewey
1991) i.e.~that both these neutron stars were formed from a massive
binary system, the ages of the radio pulsars in this case differ by
over 100 million years.  The similar ages of SGR~1900+14 and
PSR~J1907+0918 (see below) would require the progenitor stars to have
essentially identical main sequence lifetimes and hence masses. For
example, to produce two neutron stars that differ in age by only
$10^4$ yr, we estimate from eq.~1.3.9 of Shapiro \& Teukolsky (1983)
\nocite{st83} that main sequence stars of the order of 6 M$_{\odot}$
would differ in their absolute initial mass by only 0.05 $M_{\odot}$,
somewhat unlikely given the mass ratio distributions of binary stars
(see e.g.~Dewey \& Cordes 1987 \nocite{dc87} and references
therein). Both on statistical and evolutionary grounds, we
conclude that SGR~1900+14 and PSR~J1907+0918 did not share a common origin.

\subsection{Which neutron star is associated with G42.8+0.6?}

As demonstrated in \S \ref{sec:1907}, J1907+0918 is clearly a young
pulsar. Given that plausible cases for an association with a supernova
remnant can be made for several young radio pulsars with similar
characteristic ages to J1907+0918, it is appropriate to revisit the
case for the association between SGR~1900+14 and the supernova remnant
G42.8+0.6 following our discovery of PSR~J1907+0918. There are three
possibilities to be considered: (1) SGR~1900+14 is the neutron star
produced in the supernova explosion that produced G42.8+0.6; (2)
PSR~J1907+0918 is associated with G42.8+0.6; (3) neither of
these neutron stars are associated with G42.8+0.6. 

In order to make a good case for {\it any} neutron star-supernova
remnant association, the following criteria should be satisfied (see
\nocite{kas96} Kaspi 1996): (a) the distances to both objects should
agree; (b) the ages of both objects should agree; (c) the implied
transverse velocity, based on the neutron star offset from the remnant
center and the age, should be reasonable. We now review the current
evidence for both SGR~1900+14 and PSR~J1907+0918 in connection with
what is known about G42.8+0.6.

(a) Distance estimates: PSR~J1907+0918 is estimated to lie at 7.7 kpc
based on its dispersion measure and assuming the Taylor \& Cordes
(1993) Galactic electron density model. The statistical uncertainty in
this model is at least 25\%. Hurley et al.~(1999b) \nocite{hlk+99} estimate the
distance to SGR~1900+14 to be 5.7 kpc based on a spectral analysis of
{\it ASCA} data, although no estimate of the uncertainty in this
measurement is quoted.  The distance to G42.8+0.6 is commonly taken in
the literature to be 5 kpc (see e.g.~Vasisht et al.~1994).  Whilst
this appears to be in agreement with SGR~1900+14, it should be stated
that the latter distance was derived using the notoriously unreliable
$\Sigma$--$D$ relationship. A more recent $\Sigma$--$D$ study places
G42.8+0.6 at $10\pm3$ kpc \nocite{cb98} (Case \& Bhattacharya
1998). Given the considerable uncertainties associated with this
technique, it would obviously be premature to rule out an association
between G42.8+0.6 and either of the two neutron stars in question.  In
this regard, we note that the recent discovery by Vrba et al.~(2000)
\nocite{vhl+00} of a massive star cluster only 12 arcsec from
SGR~1900+14 suggests that it may have been formed in this cluster
rather than G42.8+0.6. Vrba et al.~estimate the star cluster to lie 
at 14.5 kpc.

(b) Age estimates: The ages of any non-historical supernova remnants
are strongly coupled with their distances since absolute remnant
sizes, along with assumptions about the expansion velocities are
required to constrain the ages. Hence the age of G42.8+0.6 is also
subject to considerable uncertainty. Vasisht et al.~(1994) quote an
age of $10^4$ yr but, given the above range of distance estimates,
this could easily be uncertain by factors of a few. For SGR~1900+14,
the traditional assumptions about dipolar spin-down are thought not 
to apply and the age is quite uncertain with current estimates of
$\sim10^4$ yr (see e.g.~Kouveliotou et al.~1999). For PSR~J1907+0918, the
38-kyr characteristic age is probably indicative of its true age. This
is somewhat model dependent since the age would be reduced if e.g.~the 
birth spin period of PSR~J1907+0918 was close to its current value or even
increased if the neutron star braking is less than that expected from
pure magnetic dipole braking (see e.g.~Manchester \& Taylor 1977). In summary,
based on currently-available evidence we conclude that both neutron
stars appear to be young enough to be considered as plausible
candidates for an association with G42.8+0.6.

(c) Transverse speed estimates: Both neutron stars lie about 20 arcmin
from the center of G42.8+0.6 which implies a transverse velocity of
$4000\,\,D_7/t_4$ km s$^{-1}$ to carry either of them to their present
position. Here $D_7$ is the distance in units of 7 kpc and $t_4$ is
the age in units of $10^4$ yr. Although the exact values of $D_7$ and
$t_4$ are highly uncertain, it is unlikely that they are such that the
required velocity estimate is below 1000 km s$^{-1}$. For either of
the neutron stars, then, the implied transverse velocities would place
them at the far extremes of the presently-observed distribution
\nocite{hla93} (Harrison, Lyne \& Anderson 1993). To ultimately test
for an association between PSR~J1907+0918 and G42.8+0.6, and constrain
the age of the pulsar, a proper motion measurement is required. The
predicted pulsar proper motion is $120/t_4$ mas yr$^{-1}$. 
Future VLBI proper motion measurements of PSR~J1907+0918, perhaps
using Arecibo-Effelsberg-GBT are highly desirable.

To summarize, based on the currently-available information, we
conclude that the proposed association between G42.8+0.6 and
SGR~1900+14 is, contrary to frequent claims in the literature, far
from secure since there is no reason against arguing equally strongly
in favor of PSR~J1907+0918 as being the neutron star produced in the
supernova explosion rather than SGR~1900+14. Indeed, the additional possibility
that neither of these young neutron stars is associated with G42.8+0.6
remains attractive at this stage! As noted by Gaensler
\nocite{gae00} (2000), large positional offsets between neutron stars
and supernova remnants are more likely a result of random
line-of-sight alignments rather than genuinely associated
high-velocity neutron stars. This may well be the case here and
further observations (e.g.~deeper multi-wavelength maps of the region
and proper-motion measurements) are clearly desirable to help resolve
this most perplexing situation.

\acknowledgments
We are grateful to Chryssa Kouveliotou for suggesting the radio search
of SGR~1900+14 and for releasing the source coordinates in advance of
publication. We thank Alex Wolszczan for making the PSPM available to
us and Don Backer for providing access to the ABPP. Without either of
these instruments, the observations presented here would not have been
possible. We would also like to thank Fernando Camilo and Ingrid Stairs 
for helpful discussions concerning Arecibo timing observations, as well 
as Maura McLaughlin, Jim Cordes, Fernando Camilo, Chris Salter, Bryan 
Gaensler, Chryssa Kouveliotou, Pete Woods, and the referee, Vicky
Kaspi, for a number of useful comments 
on earlier versions of this manuscript. Arecibo Observatory is run by 
Cornell University under contract with the National Science Foundation.

\bibliographystyle{apj1d}

\begin{deluxetable}{lr}
\tablewidth{0pt}
\tablecaption{Observed and derived parameters for PSR~J1907+0918}
\tablehead{Parameter & Value}
\startdata
Right Ascension (J2000)            & $\rm 19^{\rm h} 07^{\rm m}22\fs 441(4)$ \\
Declination     (J2000)            & $\rm +09^\circ  18'      30\farcs 76(4)$\\
Barycentric Period (s)             & 0.2261071099878(6) \\
Period derivative ($10^{-15}$)     & 94.2955(4) \\
Epoch (MJD)                        & 51319 \\
Dispersion Measure (cm$^{-3}$ pc)  & 357.9(1)\\
Timing data span (MJD)             & 51257--51540\\
\hline
Flux density at 0.4 GHz (mJy)      & 0.4(2) \\
Flux density at 1.4 GHz (mJy)      & 0.3(1) \\
Flux density at 2.4 GHz (mJy)      & 0.24(2) \\
Flux density at 5.0 GHz (mJy)      & 0.18(9) \\
Mean spectral index                & --0.3(2) \\
\hline
Distance\tablenotemark{a}\, (kpc)                     & 7.7 \\
Dipole magnetic field strength\tablenotemark{b}\, $B$ ($10^{12}$ G)&4.7\\
Characteristic age\tablenotemark{b}\, $\tau$ (kyr)           & 38\\
Spin-down energy loss rate\tablenotemark{c}\, $\dot{E}$ ($10^{35}$
erg s$^{-1}$) & 3.2 \\
\enddata
\label{tab:1907}
\tablecomments{Figures in parentheses represent $1\,\sigma$
uncertainties in least-significant digits quoted. }
\tablenotetext{a}{Calculated using the Taylor \& Cordes (1993)
\nocite{tc93} Galactic electron density model}
\tablenotetext{b}{Calculated using the standard
magnetic dipole formulae viz: $B=3.2\times10^{19} \sqrt{P\dot{P}}$ Gauss;
$\tau=P/2\dot{P}$ (Manchester \& Taylor 1977)}
\tablenotetext{c}{Calculated assuming rigid-body rotation for
a neutron star moment of inertia of $10^{45}$ gm cm$^2$
(Manchester \& Taylor 1977)}
\end{deluxetable}

\end{document}